\documentclass[pss]{wiley2sp} % provides pss two-column style
\usepackage{amsmath}
\begin{document}

% Title of the article
\title{Odd-even effect of quantum discord and quantum phase transitions in the S=1/2 spin ladder with ring exchange}

% Abbreviated title for the page headers
\titlerunning{Odd-even effect and quantum phase transitions}

% Authors

\author{%
  H.L.~Huang\textsuperscript{\textsf{\bfseries 1}},
  Z.Y.~Sun\textsuperscript{\Ast,\textsf{\bfseries 1}},
  B.~Wang\textsuperscript{\textsf{\bfseries 2}}}  
\authorrunning{Sun et al.}

%E-mail-address of corresponding author
\mail{e-mail
  \textsf{sunzhaoyu2012@gmail.com}}

% author's affiliations/addresses
\institute{%
  \textsuperscript{1}\,School of Electrical and Electronic Engineering, Wuhan Polytechnic University, Wuhan 430000, China\\
  \textsuperscript{2}\,Department of Physics, Beijing Normal University, Beijing 100875, China}

\received{XXXX, revised XXXX, accepted XXXX} % do not change, will be filled in by the publisher
\published{XXXX} % do not change, will be filled in by the publisher

% Please select about four verbal keywords for your manuscript.

\abstract{%
% This is a macro for the typesetting of two-column text in an
% abstract. It will typeset the two arguments in \abstcol{}{} as the
% left and right column inside the abstract box. At the
% columnbreak there will be always a columnbreak (\par), so both
% columns start with a new paragraph. No automatic column height
% balancing is done.
%
% If used with a \titlefigure it will silently output both
% parameters as consecutive paragraphs.
%
% The macro is defined exclusively inside the argument of \abstract{};
% if used outside it will raise an error.
%
% Usage: \abstcol{<left column>}{<right column>}
\abstcol{%
The singularity of quantum correlations, such as quantum entanglement(QE) and quantum discord(QD), has been widely regarded as a valuable indicator for quantum phase transition(QPT) in low-dimensional quantum systems. In this paper, for an $L\times2$ spin ladder system 
 with ring exchange, we find that the singularity of QD  }{%
(or QE) could not indicate the critical points of the system. Instead, the QD shows a novel odd-even effect in some phases, which can be used to detect the phase boundary points. The size effect is related to the symmetry breaking of the ladder.

 }}

\maketitle

\section{Introduction}

The relation between quantum correlation and quantum phase transition
(QPT) has been extensively investigated in many-body quantum physics.\cite{QE1,QE3,QE4,QPT_QE1,QPT_QE2_extr,QE_QPT_extr1,QPT_QE4,re_examin_CC-1,Hubbard_entropy-1,QD_MPS-1,Quantum_discord_and_quantum_phase_transition_in_spin_chains,Classical_correlation_and_quantum_discord_in_critical_systems,QE_extr}
It is first found by Osterloh \textit{et al.} that the quantum entanglement
(QE)---a kind of quantum correlation---shows a singular point in the
vicinity of the QPT point of the transverse-field Ising model.\cite{QE4}
Since then, the ability of QE in detecting QPT has been investigated
in many quantum systems.\cite{QPT_QE1,QPT_QE2_extr,QE_QPT_extr1,QPT_QE4,re_examin_CC-1,Hubbard_entropy-1,QE_extr}
However, QE is not the only nature of quantum correlation, thus Olliver
\textit{et al.} \cite{Quantum_Discord_A_Measure_of_the_Quantumness_of_Correlations}
proposes to use the quantum discord (QD) to quantify all the quantumness
of correlation present in a quantum state.\cite{QD_MPS-1,Quantum_Discord_A_Measure_of_the_Quantumness_of_Correlations,Quantum_discord_and_quantum_phase_transition_in_spin_chains,Classical_correlation_and_quantum_discord_in_critical_systems,Unified_View_of_Quantum_and_Classical_Correlations,Quantum_discord_for_two_qubit_systems,Quantum_discord_for_two_qubit_X_states,Classical_Correlations_and_Entanglement_in_Quantum_Measurements}  

Just like QE, QD has also been used to study the QPTs in several one-dimensional
quantum spin chains, such as the transverse Ising chain and an anti-ferromagnetic
XXZ chain.\cite{QD_MPS-1,Quantum_discord_and_quantum_phase_transition_in_spin_chains,Classical_correlation_and_quantum_discord_in_critical_systems}
In most cases, these two measures of quantum correlation shows similar
behavior in the phase transition. For example, a discontinuity of
QE or QD can be used to detect a first-order phase transition point,
while a discontinuity or a divergence in the first derivative of QE
or QD would be related to a second-order phase transition point (or
just called the critical point).\cite{Quantum_discord_and_quantum_phase_transition_in_spin_chains}
Though QD and QE have a similar capacity in detecting QPT, they have
a fundamental difference, that is, QD can survive in separable states.\cite{Unified_View_of_Quantum_and_Classical_Correlations}
As a result, QD can capture the signal of QPT in \textit{separable
states} while QE cannot. For example, in an infinite XYZ model, the
two-spin state of the system is \textit{separable} thus pairwise QE
vanishes, however, QD survives and indicates the QPT point of the
system very well.\cite{QD_MPS-1}

In addition to these simple models, recently novel quantum phases
in complex spin systems have been studied extensively, such as models
with multiple-spin exchange interaction.\cite{ladder1}
It's believed that multiple-spin exchange interaction plays an important
role in understanding the magnetic properties in several materials
such as two-dimensional compound $La_{2}CuO_{4}$.\cite{lacuo} Among
these models, the $S=1/2$ spin ladder with four-spin ring exchange(the topology is shown in Fig. 1)
has attracted special attention.\cite{ladder1-1,ladder3,ladder3-1,ladder3-2,ladder3-3,ladder4}
Firstly, the two-leg ladder can be regarded as an intermediate topology
between one- and two-dimensional lattices, thus can give insight into
the behavior of two-dimensional systems. Secondly, due to the four-spin
ring exchange interaction, the system has a very rich phase diagram.
Using various correlation functions as order parameters, L$\ddot{a}$uchli
et al. have found six phases, and the phase diagram of the system
is illustrated in Fig. 2. The non-symmetry-breaking regions contains four
phases, i.e., a ferromagnetic (FM) phase,
a rung singlet (RS) phase, a dominant vector chirality
(VC) phase, and a dominant collinear spin correlation (CSC) phase, while the 
symmetry-breaking regions contains two 
phases, i.e., 
a staggered dimers (SD) phase and a scalar
chirality (SC) phase.\cite{ladder1,ladder3-2} 
We mark the two first-order QPT points as $\theta_{1}$ and $\theta_{2}$, the three critical points as $\theta_{3}$, $\theta_{4}$ and $\theta_{5}$, and the crossover point as $\theta_{6}$.

Recently the QPTs of the ladder have been studied through quantum information theory.\cite{ladder1-1,ladder4,Sunpssb} The first-order QPT points $\theta_1$ and $\theta_2$ are identified ambitiously by the sudden change of the entanglement concurrence.\cite{ladder1-1}   The critical point $\theta_4$, which is a highly symmetric point of the system, is identified by 
the size-independent extremal point of entanglement entropy.\cite{ladder1-1} For the crossover point $\theta_6$, recently we have provided a very effective approach to identify its location by analyzing the first-excited state of the system.\cite{Sunpssb} Thus, in this paper, we will just pay our attention to the other two critical points
$\theta_3$ and $\theta_5$.

To investigate the entanglement properties of
the ladder, Song \textit{et al.} have calculated the entanglement
concurrence \cite{def_CC} for several two-spin subsystems and the
entanglement entropy\cite{Hubbard_entropy-1} for different block
geometry, and the relation between the QE and the phase diagram is
discussed in depth.\cite{ladder1-1} However, neither the concurrence nor the entanglement entropy shows any singularity
or extreme point at the phase boundary ($\theta_{5}$)  between the
SC phase and the VC phase. Especially, in most areas of the SC phase
and VC phase, the two-spin states of different subsystems (two spins
on the rung, leg, and diagonal bond) in the ladder are all found to
be \textit{separable states}, thus the concurrence vanishes and shows
no signal for $\theta_{5}$ at all. The situation is somewhat similar
to the above mentioned infinite XYZ model. Thus, it is natural to
ask: would the QD capture the signal of the phase transition in these
\textit{separable states}? Moreover, as far as we know, the existing
works on QD are mainly limited to very simple models, and the connections
between QD and novel phases in complex systems are still less well
understood. Thus, it would be valuable to study the properties of
the QD in the ladder.

In this paper, 
firstly we give a brief introduction of QD in Sec. 2,
then we describe the Hamiltonian of the ladder 
and some numerical details in Sec. 3.
The main results are shown in Sec. 4 and some discussions are given
in Sec. 5.

\begin{figure}
\includegraphics[scale=0.35]{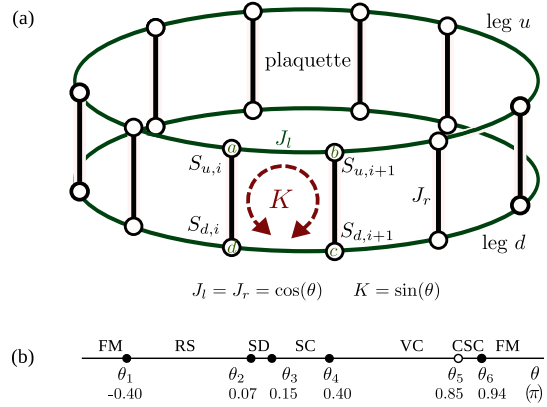}\caption{\label{fig:topo}(Color online) The topological structure of the two-leg
ladder.}
\end{figure}

\section{Quantum discord}

QD aims at characterizing
all the quantumness in a quantum state.\cite{Quantum_Discord_A_Measure_of_the_Quantumness_of_Correlations,Unified_View_of_Quantum_and_Classical_Correlations}
Its definition is based on the two quantum versions of the classical
mutual information.\cite{CIT} For a classical system (or a state)
$AB$, the total correlation between the subsystems $A$ and $B$
can be expressed as $I_{A,B}=H_{{A}}+H_{{B}}-H_{{AB}}$
or $J_{A,B}=H_{{A}}-H_{{A}|{B}}$, where
$H_{{A}}$ ($H_{{B}}$, $H_{{AB}}$) is the Shannon
entropy, and $H_{{A}|{B}}$ is the conditional entropy.
One can prove that $I_{A,B}$ and $J_{A,B}$ are
equal to each other. Now let's extend the definition of $I_{A,B}$
and $J_{A,B}$ to a quantum state described by the density
matrix $\rho_{{AB}}$. For $I_{A,B}$, by replacing
the Shannon entropy with the von Neumann entropy $S(\rho)$, one can
easily obtain the quantum mutual information $\mathcal{I}(\rho_{{AB}})$
as\cite{Quantum_Discord_A_Measure_of_the_Quantumness_of_Correlations,Quantum_discord_and_quantum_phase_transition_in_spin_chains,Classical_correlation_and_quantum_discord_in_critical_systems,Quantum_discord_for_two_qubit_systems,Quantum_discord_for_two_qubit_X_states} 

\begin{equation}
\mathcal{I}(\rho_{{AB}})=S(\rho_{{A}})+S(\rho_{{B}})-S(\rho_{{AB}}).
\end{equation}

The quantum generalization of $J_{A,B}$ is slightly difficult
because of the conditional entropy $H_{{A}|{B}}$, thus,
firstly the quantum conditional entropy $S(\rho|\{B_{k}\})$---the
quantum version of $H_{{A}|{B}}$---has to be defined,
where $\{B_{k}\}$ is just a complete set of projectors.\cite{Quantum_discord_for_two_qubit_systems,Quantum_discord_for_two_qubit_X_states}
Then a variant of quantum mutual information can be defined as $\mathcal{J}_{\{B_{k}\}}(\rho_{{AB}})=S(\rho_{{A}})-S(\rho|\{B_{k}\})$.
Following Ref. \cite{Quantum_Discord_A_Measure_of_the_Quantumness_of_Correlations},
the quantity 
\begin{equation}
\mathcal{J}(\rho_{{AB}}):=\sup_{\{B_{k}\}}\mathcal{J}_{\{B_{k}\}}(\rho_{{AB}})
\end{equation}
 is interpreted as a measure of classical correlation. Finally, QD
is just defined as the difference between the quantum mutual information
$\mathcal{I}(\rho_{{AB}})$ and the classical correlation $\mathcal{J}(\rho_{{AB}})$:
\begin{equation}
\mathcal{Q}(\rho_{{AB}})=\mathcal{I}(\rho_{{AB}})-\mathcal{J}(\rho_{{AB}}).
\end{equation}

For a state containing quantum correlation, QD is generally non-zero.
However, if $\rho_{{AB}}$ reduces into a classical state,
one can prove that $\mathcal{I}(\rho_{{AB}})$ and $\mathcal{J}(\rho_{{AB}})$
would just reduce to $I_{A,B}$ and $J_{A,B}$, respectively,
thus the QD vanishes.

For a general state, the above mathematical definition of QD is somewhat complex, with a numerical optimization procedure involved. However, for two-qubit states expressed in the following form,

\begin{equation}
\rho_{{AB}}=\left(\begin{array}{cccc}
    u & & & w\\
 & x & z\\
 & z & y\\
w  &  &  & v
\end{array}\right),\label{eq:rho-1}
\end{equation}
the QD can be simply expressed as spin-spin correlation functions,
and the numerical optimization procedure can be omitted.\cite{Quantum_discord_for_two_qubit_X_states}
Because of the symmetry of the ladder considered in this paper, the two-spin reduced
density matrix takes the form in Eq. \ref{eq:rho-1}, thus our calculation follows the formula in Ref. 
\cite{Quantum_discord_for_two_qubit_X_states}.

\section{Spin ladder with ring exchange}

We consider the following SU(2) invariant Hamiltonian defined on a
$L\times2$ two-leg ladder as\cite{ladder1}

\begin{equation}
\begin{array}{cll}
\mathcal{{H}} & = & J_{r}\sum_{i=1}^{L}\hat{S}_{u,i}\cdot\hat{S}_{d,i}\\
 &  & +J_{l}\sum_{i=1}^{L}(\hat{S}_{u,i}\cdot\hat{S}_{u,i+1}+\hat{S}_{d,i}\cdot\hat{S}_{d,i+1})\\
 &  & +K\sum_{i=1}^{L}(\hat{P}_{i,i+1}+\hat{P}_{i,i+1}^{-1}),
\end{array}\label{eq:H}
\end{equation}
 where the two legs of the ladder are marked as $u$ and $d$, and
the rungs are labeled as $i=1,\cdots,L$(see Fig. 1). The first term
in the above summation describes the interaction on the rungs, the
second term is the interaction along the legs, the third term is the
four-spin exchange interaction in the plaquettes of the ladder, and
$J_{r}$, $J_{l}$ and $K$ are the corresponding coupling constants.
$\hat{S}_{u,i}$($\hat{S}_{d,i}$) is the spin-1/2 operator defined
on the $i$-th spin in leg $u$($d$). $\hat{P}_{i,i+1}$($\hat{P}_{i,i+1}^{-1}$)
is the four-spin ring operator, which rotates the four spins in the
$i$-th plaquette clockwise (counterclockwise), i.e.,\cite{ladder1-1}

\begin{equation}
\begin{array}{ccc}
\hat{P}\vert\begin{array}{cc}
a & b\\
d & c
\end{array}\rangle=\vert\begin{array}{cc}
d & a\\
c & b
\end{array}\rangle & \textrm{and} & \hat{P}^{-1}\vert\begin{array}{cc}
a & b\\
d & c
\end{array}\rangle=\vert\begin{array}{cc}
b & c\\
a & d
\end{array}\rangle\end{array}.
\end{equation}
In this paper the periodic boundary condition is imposed by setting
$\hat{S}_{u,1}=\hat{S}_{u,L}$ and $\hat{S}_{d,1}=\hat{S}_{d,L}$.
In addition, following Ref. \cite{ladder1}, we set $J_{r}=J_{l}=\cos(\theta)$
and $K=\sin(\theta)$. The phase diagram is shown in Fig. 2.

We will investigate the behavior of quantum correlation in the vicinity of $\theta_3$ and $\theta_5$. For finite-size ladders with $L=6,8,10,12$, firstly, we use the exact
diagonalization method to obtain the ground-state wavefunction. Then we calculate the reduced density
matrix of the concerned subsystem by tracing out all other degrees
of freedom. Finally, QD can be obtained analytically according to Ref. \cite{Quantum_discord_for_two_qubit_X_states}.

We have calculated the QD for the two spins on the rung(sites $S_{u,i}$
and $S_{d,i}$), leg(sites $S_{u,i}$ and $S_{u,i+1}$), and diagonal
bond(sites $S_{u,i}$ and $S_{d,i+1}$) in the ladder, and the most
interesting results are the behavior of QD on the rung. Thus, firstly,
we limit ourselves to the two spins on the rung, and discussions about
other two-spin subsystems will be given in Sec. 5.

\begin{figure*}
\includegraphics[scale=0.5]{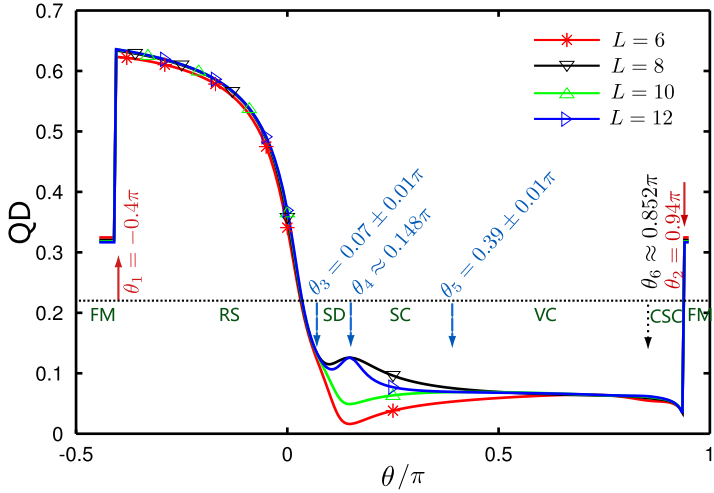}\caption{\label{fig:PHASE} (Color online) Ground-state QD of two spins on
the $i$-th rung as a function of $\theta$ in $L\times2$ spin ladders.
The phase boundaries are marked as $\theta_{i}$. Arrows with solid
line, dash line and dot line denote first-order QPT points ($\theta_{1}$
and $\theta_{2}$), second-order QPT points ($\theta_{3}$, $\theta_{4}$
and $\theta_{5}$), and the crossover point ($\theta_{6}$), respectively.
The values of $\theta_{i}$ from Ref. \cite{ladder1}
are shown in the figure.}
\end{figure*}

\section{Odd-even effect between phase boundaries $\theta_{3}$ and $\theta_{5}$}

The QD between the two spins on the $i$-th rung of $L\times2$ ladders
with $L=6,8,10,12$ is shown in Fig. 2. 

One can see that at the QPT points $\theta_1$, $\theta_2$ and $\theta_4$, the behavior of QD is similar to the entanglement reported in previous studies,\cite{ladder1-1} that is, QD is discontinuous at the first-oder QPT points  $\theta_1$ and $\theta_2$, and  shows a size-independent extremal point at the highly symmetric point $\theta_4$. 

Now we try to find the signal for  $\theta_3$ and $\theta_5$. From Fig. 2, it is expected that QD would not show any singularity or extremal point at these two points when $L$ is increased, thus we need to change our strategy. We find that the finite-size effect in the symmetry-breaking
regions ($\theta\in[\theta_{3},\theta_{5}]$) is obviously larger than that in the non-symmetry-breaking
regions ($\theta\notin[\theta_{3},\theta_{5}]$), which may be useful to detect $\theta_{3}$ and $\theta_{5}$.

In the non-symmetry-breaking regions, the size effect is very small.
For the FM phase, for example, in
the rigorous bounds $-\pi<\theta<-\frac{\pi}{2}$, it is clearly that
the ferromagnetic state minimizes the energy on each plaquette locally,
thus the finite-size effect is relatively small. For the RS phase,
let's choose the point
$\theta=0$, i.e., the spin ladder with only anti-ferromagnetic bilinear
couplings. It's well-known that the ground state can
be well approximated by the product of local rung singlets, thus the
size effect should be small in this phase. In the main region of the
VC phase and the CSC phase, QD also shows very little size dependence.

In the symmetry-breaking regions, peculiar size effect is observed,
that is, the behavior of QD with $L=4i$
is notably different from that with $L=4i+2$. The physical origin of the peculiar size
effect can be understood as follows. 
In the SD and SC phases, the translational symmetry is broken and the ``unit'' of the ground state consists
of two rungs, thus, the number of  ``unit''  is \textit{even}   for $L=4i$ and \textit{odd}   for $L=4i+2$. Then, the wavefunction and quantities such as QD would show an odd-even dependence upon the number of  ``unit'' of the ground state as follows:

\begin{equation}
\begin{array}{c}
\mbox{\ensuremath{\vert\Phi_{g}\rangle=\{\begin{array}{cl}
\vert\psi_{L}\rangle & \textrm{for}\, L=4i\\
\vert\varphi_{L}\rangle & \textrm{for}\, L=4i+2
\end{array}}}\end{array},\label{eq:selection}
\end{equation}
 and

\begin{equation}
\mathcal{Q}_{g}=\{\begin{array}{cl}
\mathcal{Q}_{\psi_{L}} & \textrm{for}\, L=4i\\
\mathcal{Q}_{\varphi_{L}} & \textrm{for}\, L=4i+2
\end{array}.\label{eq:selection_QD}
\end{equation}

In order to see the size effect more clearly,
for different phases 
we numerically calculate the QD for the two lowest-lying energy
eigenstates, and several typical results are shown
in Fig. 3. For a given $\theta$, it is obvious that these ($L$,$\mathcal{Q}$)
points belong to two curves, thus we label one of the two curves with
squares and dot line ($\psi_{L}$) and the other with circles and
dot line ($\varphi_{L}$). In addition, for the ground state, we mark
the corresponding points ($L$,$\mathcal{Q}_{g}$) by red triangles.
In non-symmetry-breaking phases(see Fig. 3 (a)(e)(f)), one finds
that the ground state is always $\vert\varphi_{L}\rangle$, while
$\vert\psi_{L}\rangle$ is always an excited state. In symmetry-breaking
phases(see Fig. 3 (b)(c)(d)), however, with the increase of $L$,
($L$,$\mathcal{Q}_{g}$) jumps between the two curves, which means
these two states take turns to serve as the ground state as the change
of $L$, just as described by Eqs. \ref{eq:selection} and \ref{eq:selection_QD}.

This jumping behavior in Fig. 3 can be used to detect the phase boundary
points $\theta_{3}$ and $\theta_{5}$. However, much manual work would be required to obtain Fig. 3 for all the $\theta$.
Based on Fig. 2, there is an alternative approach to locate $\theta_{3}$ and $\theta_{5}$ roughly but very efficiently.
First, one defines a parameter $q_{L}(\theta):=\vert\mathcal{Q}_{L}(\theta)-\mathcal{Q}_{L-2}(\theta)\vert$.
From Fig. 2 one sees that with the increase of $L$, $q_{L}(\theta)$
vanishes very rapidly in the RS and VC phases. For a given $\theta$, with $L$ large enough, suppose $q_{L}(\theta)$ is smaller
than some threshold value $\delta$, we say that the system is
in the non-symmetry-breaking region, thus $\theta_{3}$ and $\theta_{5}$ can be identified approximately.
In fact, when using $L=12$ and $\delta\sim10^{-4}$, $\theta_{3}$ and $\theta_{5}$ are found to be $0.07\pi$
and $0.39\pi$, respectively, which is consistent with previous studies.\cite{ladder1}

From Fig. 3 one can see that the odd-even behavior decreases with increasing
system size. Therefore, one might wonder, if this criterion is useful
for large systems. Keep in mind is that, for a given $\theta$, if
the odd-even behavior is observed for a small $L$, the system
should be in the symmetry-broken phase, even if the behavior becomes
blurred when the size of the system is large. Thus, for a finite $L$, the
odd-even behavior would be of practical use to detect, at least,
the ``inner boundary'' of the symmetry-broken phases. In other words, the
above values can always be regarded as an upper (lower) bound for
the critical values $\theta_{3}$ ($\theta_{5}$) of the ladder systems.

\begin{figure}
\includegraphics[scale=0.42]{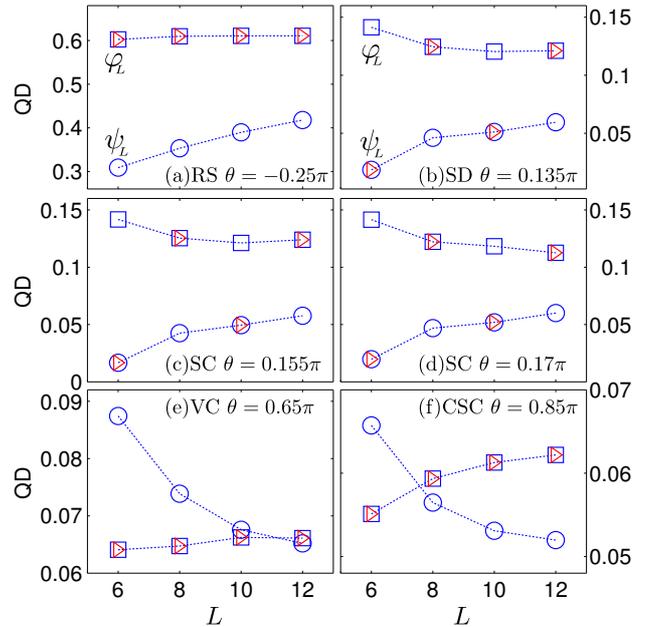}\caption{\label{fig:Q1Q2_vs_PB} (Color online) The QD of the two lowest-lying
energy eigenstates of the system for several $\theta$. The red triangle
denotes the QD of the ground state. In the SD and SC phases, the ground-state
QD jumps between two curves, which indicates that two states $\vert\psi_{L}\rangle$
and $\vert\varphi_{L}\rangle$ take turns to serve as the ground state.}
\end{figure}

\section{Discussions and summaries}

In many systems, the singularity of ground-state wavefunction is widely used to detect the QPTs (the singularity of quantum entanglement and quantum discord is also inherited from the wavefunction). However, in this paper, in a finite-size ladder with ring exchange, the wavefunction keeps analytical at $\theta_3$ and $\theta_5$. Fortunately, we find that the odd-even effect induced by the symmetry breaking of the wavefunction can be used to detect the QPT point of the model. 

The method may be a little rough compared with other large-scale numerical methods dealing with the infinite-size systems.\cite{ladder4} However, it captures the approximate locations of the QPTs with the negligible amount of calculation (the results are obtained using a standard PC in about two days). Note that a valuable estimation of the QPT point would always save much time before large-scale numerical calculations are adopted to figure out the accurate location of the QPTs. In addition, as
we have shown, the size dependence of QD provides a special perspective
to understand the physical nature of the symmetry-breaking phases.

As QD and the concurrence are closely related quantities,
we'd like to compare the behavior of QD and the concurrence in the system.
Both of them describe the quantum correlation between the two spins in the system.
QD is meaningful
in the whole parameter space: it can be used for 
both entangled states and separable states. The concurrence, as has
been mentioned, would vanish in a separable two-spin state even if
the whole system is in an entangled state.\cite{QD_MPS-1}
That's why the QD can be used to detect $\theta_5$ while the concurrence fails.\cite{ladder1-1}
Thus, when
used as phase transition indicator, QD has a broader scope of application than the concurrence.

Our studies are limited to the QD of the rung. In fact, for other two-site subsystems, such as two spins on one leg or
the diagonal bond in the ladder, the QD has also been calculated(the
result has not been shown in this paper), however, QD does not show
such obvious odd-even behaviors,
thus there is no clear signal for the boundaries $\theta_{3}$ and
$\theta_{5}$. The reason is as follows. The observed
finite-size effect always contains two sources: source (i) is just
the selection between $\vert\psi_{L}\rangle$ and $\vert\varphi_{L}\rangle$, described by Eq. \ref{eq:selection}, and
source (ii) comes from the convergence speed of the wavefunction $\vert\varphi_{L}\rangle$($\vert\psi_{L}\rangle$)
when increasing $L$. 
Source (i) just emerges in symmetry-breaking systems, while source (ii) is generally present in low-dimensional quantum systems.
To get a clear observation of
odd-even effect from source (i),  
the effect from source (ii) need
to be weak enough, that is, observables (such as QD) calculated from
$\vert\varphi_{L}\rangle$($\vert\psi_{L}\rangle$) 
need to converge very fast. Because of the special topology of the
ladder, it is expected that the speed of convergence for inter-rung
correlations would be slower than that of the intra-rung correlation.\cite{speed_converge}  
As a result, for QD of the two spins on the rung, the effect from
source (ii) is weak enough thus the effect from source (i) can be
revealed apparently, while for QD between different rungs, the effect
from source (ii) is very large thus the effect from source (i) becomes
blurred. Thus, it is expected that QD of the spins on the rung, rather
than other two-spin subsystems, is most likely to reveal the odd-even effect
from source (i) and show signals for $\theta_{3}$ and $\theta_{5}$.

The results obtained in this paper stimulate us to study the size
effect of quantum correlation in some other models with spatially
inhomogeneous phases.\cite{dimer}

\begin{acknowledgement}
This work was supported by the National Natural Science Foundation
of China (No. 11204223). This work was also supported by the Talent Scientific Research Foundation
of WHPU (No. 2012RZ09 and 2011RZ15).
\end{acknowledgement}


\begin{thebibliography}{[10]}

\bibitem{QE1}A. Einstein, B. Podolsky, and N. Rosen, Phys. Rev. \textbf{47},
777 (1935).

\bibitem{QE3} C. H. Bennett, D. P. DiVincenzo, J. A. Smolin, and
W. K. Wootters, Phys. Rev. A \textbf{54}, 3824 (1996).

\bibitem{QE4} A. Osterloh, L. Amico, G. Falci, and R. Fazio, Nature
(London) \textbf{416}, 608 (2002).

\bibitem{QPT_QE1} N. Laflorencie, E. S. S$\phi$rensen, M. S. Chang,
and I. Affleck, Phys. Rev. Lett. \textbf{96}, 100603 (2006).

\bibitem{QPT_QE2_extr} S. J. Gu, S. S. Deng, Y. Q. Li, and H. Q.
Lin, Phys. Rev. Lett. \textbf{93}, 086402 (2004).

\bibitem{QE_QPT_extr1}L.-A. Wu, M. S. Sarandy, and D. A. Lidar, Phys.
Rev. Lett. \textbf{93}, 250404 (2004).

\bibitem{QPT_QE4} O. Legeza and J. S$\acute{o}$lyom, Phys. Rev.
Lett. \textbf{96}, 116401 (2006). % two point entropy


\bibitem{re_examin_CC-1} M. F. Yang, Phys. Rev. A \textbf{71}, 030302
(2005).% reeaxmination QPT and entanglement


\bibitem{Hubbard_entropy-1}D. Larsson and H. Johannesson, Phys. Rev.
Lett. \textbf{95}, 196406 (2005).

\bibitem{QD_MPS-1}Z. Y. Sun, L. Li, K. L. Yao, G. H. Du, J. W. Liu,
B. Luo, N. Li, and H. N. Li, Phys. Rev. A \textbf{82}, 032310 (2010).

\bibitem{Quantum_Discord_A_Measure_of_the_Quantumness_of_Correlations}H.
Ollivier and W. H. Zurek, Phys. Rev. Lett. \textbf{88}, 017901 (2002);
W. H. Zurek, Rev. Mod. Phys. \textbf{75}, 715(2003).

\bibitem{Quantum_discord_and_quantum_phase_transition_in_spin_chains}R.
Dillenschneider, Phys. Rev. B \textbf{78}, 224413 (2008).

\bibitem{Classical_correlation_and_quantum_discord_in_critical_systems}M.
S. Sarandy, Phys. Rev. A \textbf{80}, 022108 (2009).

\bibitem{Unified_View_of_Quantum_and_Classical_Correlations}K. Modi,
T. Paterek, W. Son, V. Vedral, and M. Williamson, Phys. Rev. Lett.
\textbf{104}, 080501 (2010).

\bibitem{Quantum_discord_for_two_qubit_systems} S. Luo, Phys. Rev.
A \textbf{77}, 042303 (2008).

\bibitem{Quantum_discord_for_two_qubit_X_states}M. Ali, A. R. P.
Rau, and G. Alber, Phys. Rev. A \textbf{81}, 042105 (2010).

\bibitem{Classical_Correlations_and_Entanglement_in_Quantum_Measurements}V.
Vedral, Phys. Rev. Lett. \textbf{90}, 050401 (2003).

\bibitem{def_CC} W. K. Wootters, Phys. Rev. Lett. \textbf{80}, 2245
(1998). % C.C. definetion


\bibitem{CIT}C. E. Shannon and W. Weaver, The Mathematical Theory
of Communication (University of Illinois Press, Urbana, IL, 1949);
T. M. Cover and J. A. Thomas, Elements of Information Theory (John
Wiley $\&$ Sons), New York, 1991.


\bibitem{lacuo}Y. Honda, Y. Kuramoto, and T. Watanabe, Phys. Rev.
B \textbf{47}, 11329 (1993). 

\bibitem{ladder1}A. L$\ddot{a}$uchli, G. Schmid, and M. Troyer,
Phys. Rev. B \textbf{67}, 100409 (2003). 

\bibitem{ladder1-1}J. L. Song, S. J. Gu, and H. Q. Lin, Phys. Rev.
B \textbf{74}, 155119 (2006).

\bibitem{ladder3}I. Maruyama, T. Hirano, and Y. Hatsugai, Phys. Rev.
B \textbf{79}, 115107(2009).

\bibitem{ladder3-1}M. Arikawa, S. Tanaya, I. Maruyama, and Y. Hatsugai1,
Phys. Rev. B \textbf{79}, 205107 (2009).

\bibitem{ladder3-2}T. Hikihara, T. Momoi, and X. Hu, Phys. Rev. Lett.
\textbf{90}, 087204 (2003).

\bibitem{ladder3-3}I. Maruyama, T. Hirano, and Y. Hatsugai, Phys.
Rev. B \textbf{79}, 115107 (2009).

\bibitem{ladder4}S. H. Li, Q. Q. Shi, Y. H. Su, J. H. Liu, Y. W. Dai, and H. Q. Zhou, Phys. Rev. B 86, 064401 (2012).

\bibitem{Sunpssb} Z. Y. Sun, H. L. Huang, and B. Wang, Phys. Status Solidi B, doi:10.1002/pssb.201248368.

\bibitem{QE_extr}S. S. Deng, S. J. Gu, and H. Q. Lin, Phys. Rev.
B \textbf{74}, 145103 (2006).

\bibitem{speed_converge} For a general system there is no rigorous proof. Here we just provide some phenomenological discussions. For a simple one-dimensional chain, it is usually found that the magnetization on a site converges faster than the correlation function between two nearest sites. To expand to the two-leg ladder, the rung can be seen as a super-site thus the intra-rung correlation just cover one site along the chain direction, while the inter-rung correlations cover two sites. Along the chain direction, an operator with a short span usually has a faster convergence than that with a long span.


\bibitem{dimer} O. Legeza, J. S$\acute{o}$lyom, L. Tincani and R. M. Noack, Phys. Rev. Lett. \textbf{99}, 087203 (2007).



\end{thebibliography}
\end{document}